\pdfoutput=1
\documentclass[10pt,orivec]{llncs}
\usepackage[T1]{fontenc}
\usepackage{amssymb}
\usepackage{amsmath}
\usepackage{amsfonts}
\usepackage{dcolumn}
\usepackage{bm, color}
\usepackage{graphicx} 
\usepackage{diagbox} 
\usepackage{subfig}
\usepackage{caption}
\usepackage{algpseudocode}
\usepackage{algorithmicx}
\usepackage{algorithm}
\usepackage{makecell}
\usepackage{multirow}
\usepackage{hyperref}
\usepackage{smartdiagram}
\usepackage{verbatim}
\hypersetup{
    colorlinks=true,
    linkcolor=blue,
    filecolor=blue,      
    urlcolor=blue,
    citecolor=blue,
}
\usepackage{quantikz}

\usepackage{tikz}

\usetikzlibrary{positioning,patterns}
\usetikzlibrary{decorations.pathreplacing,decorations.pathmorphing}
\colorlet{colred}{red!22}
\colorlet{colcyan}{cyan!22}
\colorlet{colV}{blue!40}
\colorlet{colBorder}{gray!70}
\tikzset
  {mybox/.style=
    {rectangle,rounded corners,drop shadow,minimum height=1cm,
     minimum width=2cm,align=center,fill=#1,draw=colBorder,line width=1pt
    },
   myarrow/.style=
    {draw=#1,line width=3pt,-stealth,rounded corners
    },
   mylabel/.style={text=#1}
  }
\usepackage{array} 
\renewcommand\arraystretch{1.3} 
\usepackage{booktabs}
\usepackage{enumerate}

\usepackage{url}
\usepackage{orcidlink}

\renewcommand{\epsilon}{\varepsilon}

\renewcommand{\c}{\ensuremath{\mathcal{C}}}

\newcommand{\e}{\ensuremath{\mathcal{E}}}

 \newcommand{\tr}{{\rm tr}} 
 \renewcommand{\a}{\ensuremath{\mathcal{A}}}

\newcommand{\jg}[1]{\color{red} {JG: #1 :JG} \color{black} }

\newcommand{\mvs}{\fontfamily{mvs}\fontencoding{U}\fontseries{m}\fontshape{n}\selectfont}
\newcommand\Letter{{\mvs\char66}}

\newif\ifbadgeavailable\newif\ifbadgefunctional\newif\ifbadgereusable
\badgeavailabletrue
\badgefunctionalfalse
\badgereusablefalse
\RequirePackage{graphicx}
\usepackage[firstpageonly=true,angle=0,vpos=.147\paperheight,hpos=.393\linewidth,vanchor=t,hanchor=l]{draftwatermark}
\SetWatermarkText{%
\raisebox{-5.5cm}
{\ifbadgeavailable\includegraphics[width=11mm]{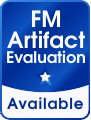}\hspace{.815\linewidth}\else\hspace{.905\linewidth}\fi%
\ifbadgefunctional\includegraphics[width=11mm]{badges/FM_2024_AE_functional}\else\ifbadgereusable\includegraphics[width=11mm]{badges/FM_2024_AE_reusable}\fi\fi}}

\begin{document}
\title{\textcolor{red}{Veri}\textcolor{green}{Q}\textcolor{blue}{R}: A \textcolor{blue}{R}obustness \textcolor{red}{Veri}fication Tool for \textcolor{green}{Q}uantum Machine Learning Models}

\author{Yanling Lin\inst{1,2}~\orcidlink{0009-0009-5492-3752}, Ji Guan\inst{2}\textsuperscript{(\Letter)}\thanks{guanj@ios.ac.cn}~\orcidlink{0000-0002-3490-0029}, Wang Fang\inst{2}~\orcidlink{0000-0001-7628-1185}, Mingsheng Ying\inst{3}~\orcidlink{0000-0003-4847-702X}, \\Zhaofeng Su\inst{1}\textsuperscript{(\Letter)}\thanks{zfsu@ustc.edu.cn}~\orcidlink{0000-0003-0021-225X}}
\authorrunning{Y. Lin et al.}



%

\newcommand{\custominstitute}[1]{%
  \parbox{0.7\textwidth}{\centering #1}
}

\institute{University of Science and Technology of China, Hefei 230026, China \and Key Laboratory of System Software (Chinese Academy of Sciences) \newline and State Key Laboratory of Computer Science, \newline Institute of Software, Chinese Academy of Sciences, \newline Beijing 100190, China \and Centre for Quantum Software and Information, University \newline of Technology Sydney, NSW 2007, Australia}


\maketitle              
\pagestyle{plain}

\begin{abstract}
Adversarial noise attacks present a significant threat to quantum machine learning (QML) models, similar to their classical counterparts. This is especially true in the current Noisy Intermediate-Scale Quantum era, where noise is unavoidable. Therefore, it is essential to ensure the robustness of QML models before their deployment. To address this challenge, we introduce \textit{VeriQR}, the first tool designed specifically for formally verifying and improving the robustness of QML models, to the best of our knowledge. This tool mimics real-world quantum hardware's noisy impacts by incorporating random noise to formally validate a QML model's robustness. \textit{VeriQR} supports exact (sound and complete) algorithms for both local and global robustness verification. For enhanced efficiency, it implements an under-approximate (complete) algorithm and a tensor network-based algorithm to verify local and global robustness, respectively. 
As a formal verification tool, \textit{VeriQR} can detect adversarial examples and utilize them for further analysis and to enhance the local robustness through adversarial training, as demonstrated by experiments on real-world quantum machine learning models. Moreover, it permits users to incorporate customized noise. Based on this feature, we assess \textit{VeriQR} using various real-world examples, and experimental outcomes confirm that the addition of specific quantum noise can enhance the global robustness of QML models. These processes are made accessible through a user-friendly graphical interface provided by \textit{VeriQR}, catering to general users without requiring a deep understanding of the counter-intuitive probabilistic nature of quantum computing.\\

The source code of VeriQR is available at \url{https://github.com/Veri-Q/VeriQR}, while the artifact for reproducing the experiments of this paper is available at \cite{VeriQR_artifact}.


\end{abstract}

\section{Introduction}
Over the last decade, machine learning (ML) has driven technological advancements in various fields. The combination of machine learning with quantum computing has given rise to a new field of research known as \emph{quantum machine learning (QML)}. In classical ML, classification models are vulnerable in adversarial scenarios \cite{chakraborty2021survey,biggio2018wild}. Specifically, the addition of intentionally crafted noises to the original data can cause classifiers to make incorrect predictions with high confidence. An illustrative example is the misclassification of a panda image as a gibbon with a confidence level exceeding 99\% after adding imperceptible noise \cite{szegedy2013intriguing}. Although studies have shown the potential superiority of quantum computers over classical counterparts in certain well-known ML tasks~\cite{biamonte2017quantum}, the presence of noise in quantum computation is inevitable due to the limitations of quantum hardware devices in the current Noisy Intermediate-Scale Quantum (NISQ) era~\cite{preskill2018quantum}, which may cause quantum learning systems to suffer from adversarial perturbations from environmental noises. 
Research on the vulnerability of QML models has garnered widespread attention \cite{lu2020quantum,du2021quantum,liu2020vulnerability,weber2021optimal,helstrom1967detection,guan2021robustness,guan2022verifying}. 
In particular, formal methods have been employed to verify the robustness of QML models against noises. Various algorithms have been developed to verify both local and global robustness, which have established a formal framework for verifying the robustness of QML models, allowing for detecting non-robust quantum states (also known as quantum adversarial examples) during the verification process. 


Numerous tools have been developed to verify the robustness of classical ML models and improve robustness through adversarial training. Notable examples include NNV~\cite{tran2020nnv}, Reluplex~\cite{katz2017reluplex}, DeepG \cite{balunovic2019certifying}, PRODeep~\cite{li2020prodeep}, VerifAI~\cite{dreossi2019verifai} and AI$^2$~\cite{gehr2018ai2}. These tools have simplified the process for users to verify the robustness of their ML models. However, understanding the counter-intuitive principles of quantum mechanics, which serve as the inherent probabilistic foundation of quantum systems, poses a distinctive challenge for the average user. Therefore, there is a requirement for automated tools in the analysis of quantum systems.

In recent years, formal methods-based tools have emerged to verify the correctness of quantum systems. For instance, the development of a specification language and an automated tool called AUTOQ enables symbolic verification of quantum circuits~\cite{chen2023autoq}. Similarly, CoqQ, integrated into the Coq proof assistant, provides a means to reason about quantum programs \cite{zhou2023coqq}. A measurement-based linear-time temporal logic (MLTL) has been proposed to formally check the quantitative properties of quantum algorithms~\cite{guan2024measurement}. Furthermore, model checkers like QMC~\cite{papanikolaouqmc} and QPMC~\cite{feng2015qpmc} have been proposed for verifying quantum programs and communication protocols.   However, to the best of our knowledge, there are currently no dedicated tools available for verifying the robustness of QML models and then improving robustness.

\begin{figure}[h]
    \vspace{-0.3cm}
    \centering
    \includegraphics[width=\linewidth]{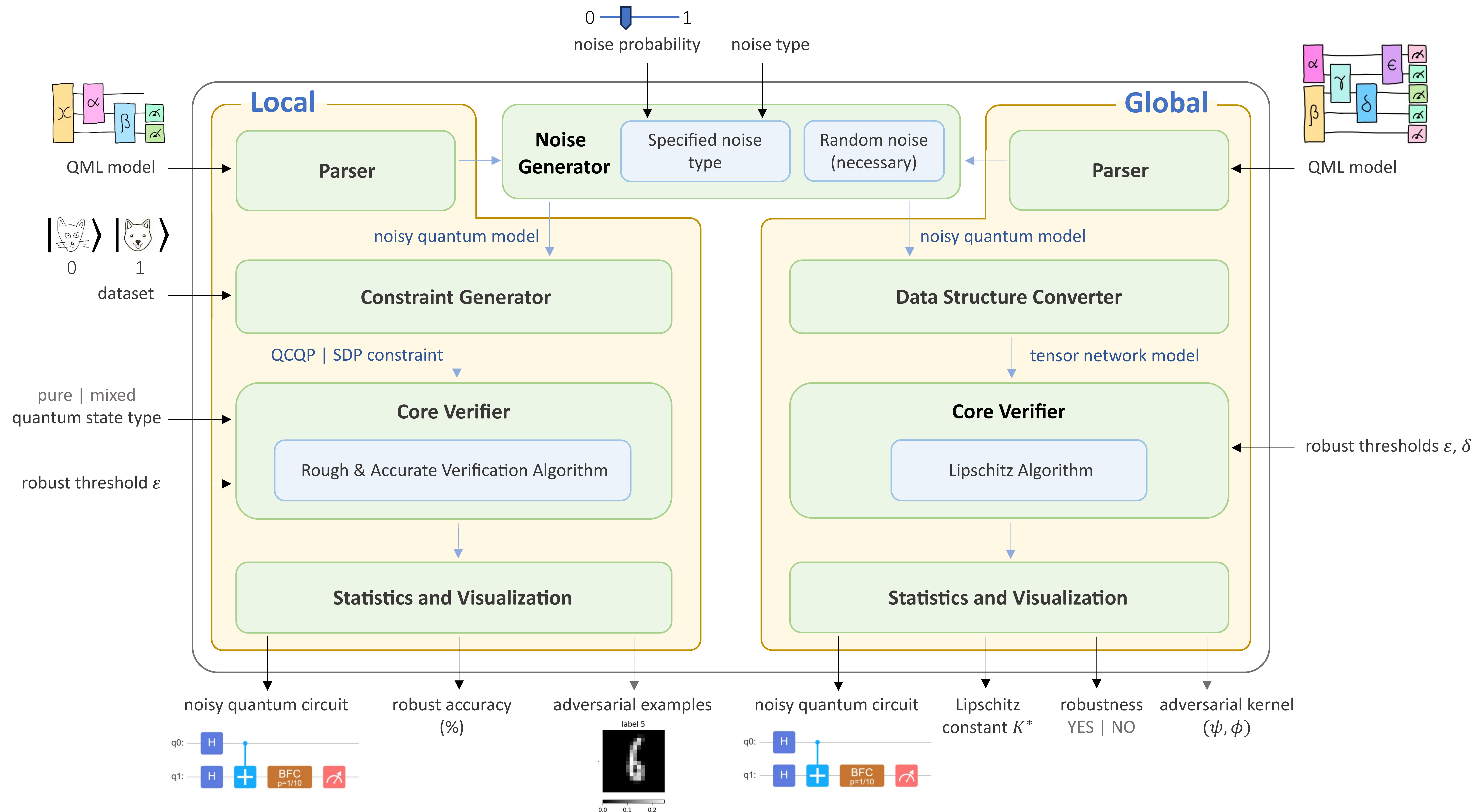}
    \caption{An overview of the  architecture of \textit{VeriQR}. }
    \label{fig:architecture}
    \vspace{-0.4cm}
\end{figure}

\paragraph{\textbf{Contributions.}} To fill the gap mentioned above, we introduce a tool named \textit{VeriQR}.  \textit{VeriQR} is built upon the aforementioned theoretical formal verification techniques~\cite{guan2021robustness,guan2022verifying} for automatically quick robustness verification of QML models and the improvement strategies for enhancing robustness. The architecture of \textit{VeriQR} is shown in Fig.~\ref{fig:architecture} and its main advantages are listed in the following. 
\begin{enumerate}
    \item For \emph{universality}, \textit{VeriQR} supports the verification of two distinct robustness properties. These are referred to as \textit{local robustness} for QML classification models and \textit{global robustness} for all existing QML models.
    \item For \emph{usability}, \textit{VeriQR} offers support for QML models that are represented in the OpenQASM 2.0 format of IBM \cite{cross2017open}. This format is widely utilized as a programming language for describing quantum circuits and algorithms.
    \item For \emph{reality}, \textit{VeriQR} formally verifies the robustness of a QML model by adding random noise to the model. This functionality enables simulations of the noisy effects of real-world quantum hardware on the robustness verification of various QML models. 
    \item For \emph{efficiency}, in addition to basic verification methods, \textit{VeriQR} incorporates various optimization techniques. These include approximation techniques for \emph{local robustness} and tensor network contractions for \emph{global robustness}, which enhance the performance of the verification process. 
    \item For \emph{local robustness enhancement}, \textit{VeriQR} can utilize the identified adversarial examples from the verification process for adversarial training, akin to traditional methods. Additionally, users have the option to introduce customized noise for \emph{improving global robustness}, as discussed in~\cite{du2021quantum,guan2022verifying,huang2023certified}. This customized noise extends beyond standard quantum noise to include user-defined quantum noise models. 
\end{enumerate}
In Section \ref{sec:evaluation}, we present experimental results demonstrating the versatility and practicality of \textit{VeriQR} in verifying and improving the robustness of different QML models in real-world scenarios. The experiments cover a range of noise types and levels, showcasing the efficacy and reliability of \textit{VeriQR}.

\section{Robustness for Quantum Machine Learning Models}
For the convenience of the reader, we briefly introduce the concepts of quantum computing used in this paper and QML models (algorithms). We then review the local and global robustness verification problems for QML models in their most basic form, which can be handled by our tool \textit{VeriQR}. For more details, please refer to ~\cite{guan2021robustness,guan2022verifying,nielsen2001quantum}. 


\subsection{Quantum Machine Learning Model}\label{subsection:QML}
A QML model $\a$ is composed of input quantum states, a quantum circuit, and a quantum measurement. 

\vspace{-0.2cm}
\paragraph{\textbf{Quantum state.}}
The input quantum state $\rho$ refers to the data that is processed by the quantum model. Mathematically, $\rho$ is a positive semi-definite matrix with a size of $2^n$-by-$2^n$, where $n$ represents the number of quantum bits (qubits). It is important to note that the quantum state $\rho$ can not only represent quantum information, such as the state of a physical Hamiltonian system, for physical computational tasks but also encode classical information, such as image data or financial data, for classical computational tasks. 

\begin{figure}[h]
    \vspace{-0.2cm}
    \centering
    \includegraphics[width=\linewidth]{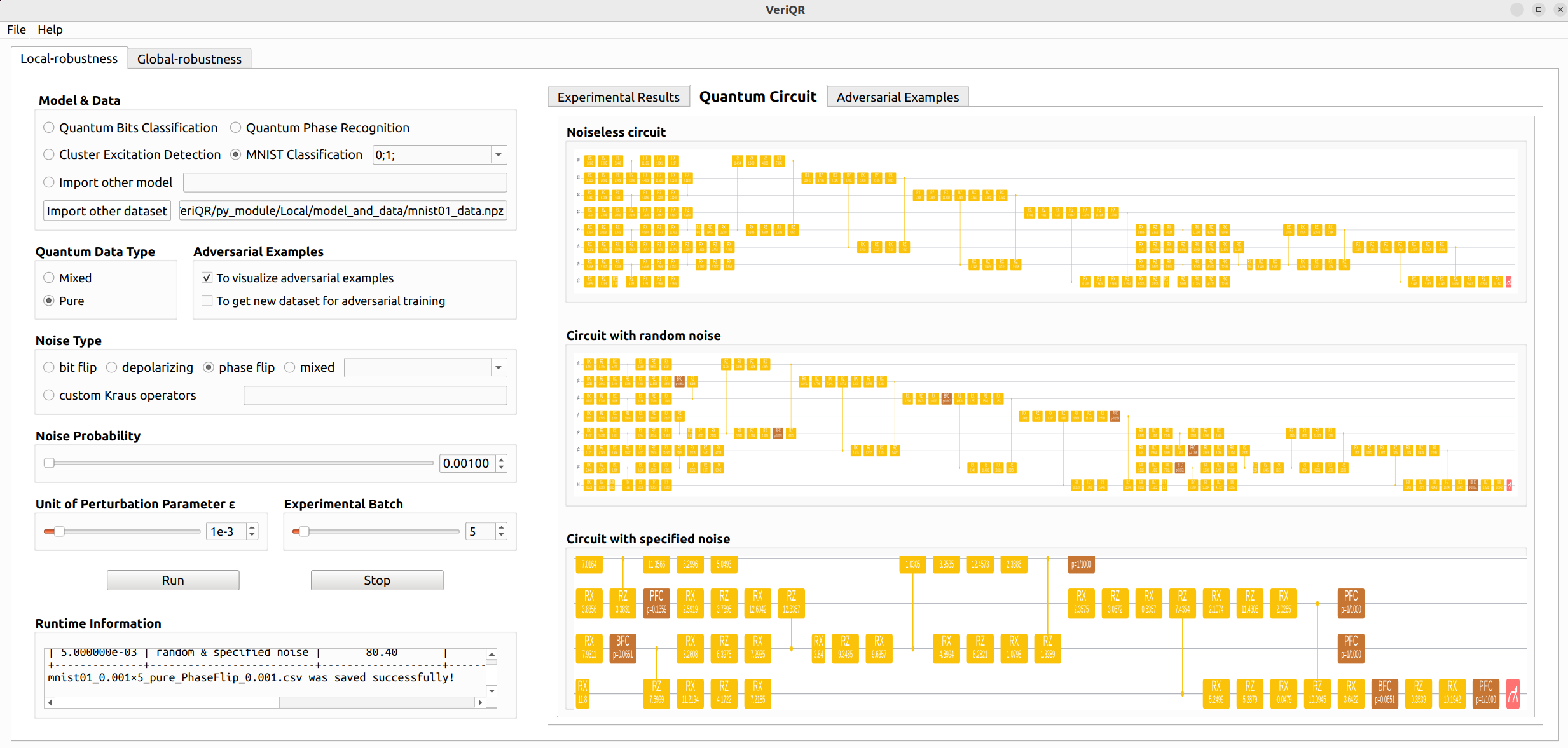}
    \caption{GUI: the main tabs for verification task and (8-qubit) quantum circuit diagrams corresponding to the QML model to be verified. In this diagram representation, yellow boxes represent 1-qubit gates, blue ones represent controlled gates (not shown in this example), brown ones represent quantum noises, and red ones represent measurements. }
    \label{fig:GUI}
    \vspace{-0.6cm}
\end{figure}

\vspace{-0.2cm}
\paragraph{\textbf{Quantum circuit.}}
The noisy quantum circuit $\e$ is used to describe the computational aspect of the QML model. A quantum circuit consists of a sequence of \textit{quantum logic gates} and \textit{quantum noises} (represented by yellow and brown boxes in Fig.~\ref{fig:GUI}, respectively).

\emph{Quantum logic gates} are the building blocks of quantum circuits and can transform a quantum state into a new quantum state, like classical logic gates are for conventional digital circuits. They are described as unitary matrices relative to some orthogonal basis. Mathematically, a gate that acts on an $n$-qubit quantum state $\rho$ is represented by a $2^n \times 2^n$ unitary matrix $U$, and its output is a evolved quantum state $\rho'=U \rho U^\dagger$, where $U^\dagger$ is the conjugate transpose of $U$. 

  \emph{Quantum noise} in quantum systems can be broadly characterized as either coherent or incoherent. Coherent noise generally originates from the noisiness of the parameters in gate operations, so it is unitary evolution (represented by a unitary matrix) and easy to simulate; incoherent noise arises from the interaction between the system and the environment and thus is usually a non-unitary evolution, which transforms the state of the quantum system from a pure state $\rho$ to a mixed state $\e(\rho)$ with 
  $
    \e(\rho)=\sum_{k}E_k \rho E_k^\dagger,
  $
  where the matrices $\{E_k\}$ with a size of $2^n$-by-$2^n$ are called \textit{Kraus operators}, satisfying the completeness conditions $\sum_{k}E_k^\dagger E_k = I$, where $I$ is the identity operator. This transformation is also known as a \textit{quantum channel}, it is a quantum operation characterized by a $2^n$-by-$2^n$ matrix. Mathematical representations of common 1-qubit quantum channels, including \textit{bit flip channel (BFC)}, \textit{phase flip channel (PFC)}, and \textit{depolarizing channel (DC)}, are described as follows: 
  \begin{equation*}
      \begin{aligned}
          \e_{\text{BFC}}(\rho)&=(1-p)I \rho I + p X \rho X\\
          \e_{\text{PFC}}(\rho)&=(1-p)I \rho I + p Z \rho Z\\
          \e_{\text{DC}}(\rho)&=(1-p)I \rho I + \frac{p}{3}(X \rho X + Y \rho Y + Z \rho Z)
      \end{aligned}
  \end{equation*}
  and 
  \[
          X=\begin{bmatrix}
          0&1\\
          1&0
          \end{bmatrix},  Y=\begin{bmatrix}
          0&-i\\
          i&0
          \end{bmatrix},  Z=\begin{bmatrix}
          1&0\\
          0&-1
          \end{bmatrix}.\]
  Here $p$ represents the likelihood of the state $\rho$ undergoing further manipulation by a quantum gate. For instance, in a bit flip channel, $p$ signifies the chance of a bit flip operation affecting the quantum state. These three categories of quantum channels are frequently encountered noise in real-world quantum hardware. In this context, $p$ serves as a measure of the noise level. A higher value of $p$ corresponds to a more pronounced alteration in the initial state $\rho$. 
  Therefore, the state of the quantum system after a noisy quantum circuit $\e$ represented by a set of matrices $\{E_k\}$ is $\e(\rho)=\sum_{k}E_k \rho E_k^\dagger$. 

\vspace{-0.2cm}
\paragraph{\textbf{Quantum measurement.}}
At the end of each quantum circuit, a quantum measurement (represented by red boxes in Fig.~\ref{fig:GUI}) is performed to extract the computational outcome, which contains classical information, from $\e(\rho)$. This information is a probability distribution over the possible outcomes of the measurement. Mathematically, a quantum measurement is modeled by a set $\{M_{c}\}_{c\in\c}$ of positive semi-definite matrices with a size of $2^n$-by-$2^n$. Here, $\c$ represents a finite set of measurement outcomes or class labels. The observation process is probabilistic: for the current state $\e(\rho)$, the measurement outcome $c\in\c$ is obtained with probability $p_{c}=\tr(M_{c}\e(\rho))$, which is the summation of diagonal entries of $M_{c}\e(\rho)$. 

In summary, a QML model, denoted as $\a=(\e,\{M_c\}_{c\in \c})$, can be viewed as a randomized mapping. For any input quantum state $\rho$, the model outputs a probability distribution $\a(\rho)=\{\tr(M_c\e(\rho))\}_{c\in\c}$.

\vspace{-0.2cm}
\paragraph{{\textbf{Pure versus mixed quantum states.}}}
It is essential to note that quantum states fall into two main categories: pure and mixed states. 
A quantum system with a known exact state $|\psi \rangle$ containing $n$ qubits is considered to be in a pure state, which can be represented by a column vector of size $2^n$ in a complex vector space. In this scenario, the density matrix (operator) representing the system is $\rho = |\psi\rangle \langle \psi|$, characterized by positive semidefinite matrices with a trace of 1. 
On the other hand, if the state of the quantum system is not precisely known, it is classified as a mixed state, which comprises an ensemble of pure states {($p_1$, $|\psi_{1}\rangle$), ($p_2$, $|\psi_{2}\rangle$), ..., ($p_n$, $|\psi_{n}\rangle$)}, denoted as $\rho = \sum_{i} p_i |\psi_{i}\rangle \langle \psi_{i}|$. This indicates that the system is in state $|\psi_{j}\rangle$ with a probability of $p_j$. A pure state $|\psi \rangle$ can be considered a special instance of the mixed state $\rho = |\psi\rangle \langle \psi|$, implying that the collection of pure quantum states is a subset of mixed quantum states. 
Pure states are mainly employed to safeguard against deliberate classical attacks (by humans) embedded in input quantum states, while mixed states are utilized in a broader array of situations, including defense against quantum noise. To accommodate different application contexts, our tool \textit{VeriQR} empowers users to choose the specific type of quantum states they wish to work with.

\subsection{Robustness Verification of QML Models}\label{Subsection:Robustness}
Similar to their classical counterparts, QML models can be delineated into two primary categories: \emph{regression models} and \emph{classification models}. Consequently, two distinct forms of robustness verification are requisite. 

\vspace{-0.2cm}
\paragraph{{\textbf{Global robustness for regression models.}}} 
A \emph{regression model} $\a$ uses the output distribution $\a(\rho)$ directly to determine the predicted value for the regression variable $\rho$. Naturally, a regression model that is robust to adversarial noise attacks should have the ability to maintain stable predictions with a certain degree of tolerance for small changes, which could induce incorrect predictions, in the initial data. In other words, it is necessary to treat all similar input states with minor differences similarly to ensure robustness for regression models, which is called \emph{global robustness}. 
\begin{problem}[Global Robustness Formal Verification]\label{Prob:Global}
Let $\a=(\e,\{M_c\}_{c\in \c})$ be a QML model, and check whether $\a$ is $(\varepsilon, \delta)$-globally robust, i.e., for any pair of quantum state $\rho$ and $\sigma$ with
$D(\rho, \sigma) \leq \varepsilon$, we have $d(\mathcal{A}(\rho), \mathcal{A}(\sigma)) \leq \delta$. If not, provide such a pair of quantum states violating the robustness. 
\end{problem}
Here $D(\cdot, \cdot)$ and $d(\cdot, \cdot)$ represent the trace distance of two density matrices and the total variance distance of the measurement outcome probability distributions on quantum states, respectively. These distances are used to quantify the similarities in input and output states, respectively. To solve the formal verification problem (Problem~\ref{Prob:Global}),  the Lipschitz constant $K^*$ of $\a$ is introduced with the fact that $\a$ is $(\varepsilon, \delta)$-globally robust, if and only if $\delta\geq \epsilon$~\cite{guan2022verifying}. Here the Lipschitz constant $K^*$ is  the smallest $K$ such that $d(\mathcal{A}(\rho), \mathcal{A}(\sigma)) \leq K D(\rho,\sigma)$ for all quantum states $\rho$ and $\sigma$. So the key is to compute $K^*$ which is done by our tool \textit{VeriQR}.

\vspace{-0.2cm}
\paragraph{{\textbf{Local robustness for classification models.}}} 
A \emph{classification model (classifier)} $\a$ utilizes the probability distribution $\a(\rho)$ to assign a class label $c\in\c$ to the input state $\rho$. The most commonly used approach is to assign the label with the highest corresponding probability in the output distribution $\{\tr(M_c\e(\rho))\}_{c\in\c}$. Naturally, a robust classifier should be able to classify all similar input states in the same class to ensure robustness, which is referred to as \emph{local robustness}. 

\begin{problem}[Local Robustness Formal Verification]\label{Prob:local}
    Let $\a=(\e,\{M_c\}_{c\in \c})$ be a QML model. Given an input state $\rho$ with label $c\in\c$, check whether $\a$ is $\epsilon$-locally robust, i.e., $\mathcal{A}(\sigma)=c$ for all $\sigma \in \mathcal{N}_\varepsilon(\rho)$, the $\varepsilon$-neighbourhood of $\rho$. If not, provide an adversarial example (counter-example) $\sigma \in \mathcal{N}_\varepsilon(\rho)$ with label $l\not=c$.
\end{problem}

Here, the $\varepsilon$-neighbourhood of $\rho$ is defined as $\mathcal{N}_\varepsilon(\rho)=\{ \sigma : \bar{F}(\rho, \sigma) \leq \varepsilon\}$, and $\bar{F}(\rho, \sigma)$ quantifies the similarity between states $\rho$ and $\sigma$ using fidelity~\cite{nielsen2001quantum}. To evaluate the $\varepsilon$-robustness of a given finite set of labeled quantum states, we assess each input example individually. Subsequently, we generate a collection of concrete adversarial examples and determine the proportion of $\varepsilon$-robust states in the dataset. This measure, referred to as the $\varepsilon$-robust accuracy of the quantum classifier $\mathcal{A}$, provides insight into its \textit{local robustness} on the dataset. 


\subsection{Challenges of Implementation}\label{Sec:challenges}
When it comes to implementing a verification tool for QML models, we encounter distinct challenges compared to dealing with classical ML models.

\vspace{-0.2cm}
\paragraph{{\textbf{Continuous state space.}}}
Quantum systems that operate within a linear space of finite dimensions possess a continuous state space. This implies that QML models must account for an infinite number of quantum states when conducting global robustness verification. In contrast, classical models mainly work with discrete input datasets that have a finite number of data. This distinction renders classical robustness verification techniques~\cite{albarghouthi2021introduction} unsuitable for quantum systems, such as the reachability method~\cite{tran2021robustness} and abstract interpretation~\cite{gehr2018ai2}.  Consequently, we develop \textit{VeriQR} as an independent tool that does not rely on any existing classical robustness tools. Instead, we implement the algorithms proposed in~\cite{guan2021robustness,guan2022verifying} to verify the robustness of QML models. 

\vspace{-0.2cm}
\paragraph{{\textbf{State explosion.}}}
The size of QML models, which is given by the dimension $2^n$, grows exponentially as the number of qubits $n$ increases. This poses challenges in terms of memory usage and runtime when performing robustness verification on large-scale systems. To address this, we employ tensor networks as an efficient data structure for storing quantum circuits, effectively optimizing memory usage. Furthermore, we utilize Google's tensor network calculator~\cite{roberts2019tensornetwork} with heuristic methods as a subroutine to enhance the efficiency of verifying global robustness (Problem~\ref{Prob:Global}). Furthermore, we have implemented the approximate verification algorithm~\cite{guan2021robustness} for local robustness verification (Problem~\ref{Prob:local}). These optimization techniques allow \textit{VeriQR} to handle robustness verification of noisy QML models with up to 20 qubits on a small service for general users (refer to Section~\ref{sec:evaluation} for experimental results). Without these optimizations, \textit{VeriQR} is only able to handle models with up to 8 qubits.

\vspace{-0.2cm}
\paragraph{{\textbf{QML benchmarks.}}}
Currently, there are only a few benchmarks available for quantum circuits (e.g., \cite{chen2022veriqbench}), and there is a lack of benchmarks specifically designed for QML models. To broaden the range of applicable scenarios, we have incorporated the use of OpenQASM 2.0 files as inputs. 
Moreover, to further enhance this capability, we have developed built-in scripts for translating QML models on several platforms (such as Huawei's MindSpore Quantum~\cite{xu2024mindspore} and Google's Cirq~\cite{Cirq}) into the OpenQASM 2.0 format. 
This enables the establishment of a unified verification benchmark framework for QML models deployed on various popular quantum platforms, including IBM's Qiskit~\cite{QiskitTextbook}, Google's TensorFlow Quantum~\cite{broughton2020tensorflow}, and others. In addition, we have visualized the framework by providing a graphical user interface (GUI) that converts inputted OpenQASM 2.0 code, used for describing quantum circuits, into visual representations (see the right side of Fig.~\ref{fig:GUI}). 

\section{Overview and Features of \textit{VeriQR}}\label{sec:overview}
\textit{VeriQR} is a graphical user interface (GUI) tool developed using C++. The decision to use C++ was influenced by the widespread use of Qt \cite{blanchette2006c++} in GUI programming. 
As shown in Fig.~\ref{fig:architecture}, \textit{VeriQR} consists of two main parts: \textit{Local robustness verification} and \textit{Global robustness verification}. 

\vspace{-0.2cm}
\paragraph{{\textbf{Inputs.}}}
To utilize \textit{VeriQR}, the user is required to import a relevant example, specifically a QML model and a dataset that contains quantum states and their corresponding ground truth labels and can be sourced from either a training or testing dataset. 
\textit{VeriQR} accepts a model in the following formats, each of which represents a quantum circuit with a measurement at the end of the circuit. 
\begin{itemize}
  \item [1. ] 
  {\it A NumPy data file (.npz format)} is utilized to package a quantum circuit, quantum measurement, and training dataset. This format is particularly beneficial for individuals who are not experts in quantum computing but have proficiency in classical formal methods and machine learning. By incorporating NumPy, \textit{VeriQR} becomes more accessible to average users without requiring extensive learning. Moreover, \textit{VeriQR} provides four popular testing examples (see the upper left corner of Fig.~\ref{fig:GUI}) of quantum classifiers in .npz, catering to beginners. 
  \item [2. ] 
  {\it An OpenQASM 2.0 file (.qasm format)} expresses the quantum circuit corresponding to the QML model to be checked. OpenQASM 2.0 is an IBM-introduced format widely adopted in the quantum computing community for constructing quantum circuits~\cite{cross2017open}. \emph{QML models trained on different quantum platforms can be converted into this format. 
 This allows for unified and reliable verification of robustness, addressing the challenge of "QML Benchmarks" discussed in Section~\ref{Sec:challenges}.}
\end{itemize}

It is important to mention that the verification of \textit{global robustness} does not require the use of the original dataset as input. Therefore, users only need to import a QML model in a .qasm file for the circuit and the measurement, without the need for additional training data. Once this step is completed, users can proceed to configure parameters for the specific case of interest. These parameters consist of the following: 
(i) the types and levels (probabilities ranging from 0 to 1) of noise:
\textit{VeriQR} inherently provides users with the option to select three standard types of noise, namely \textit{depolarizing}, \textit{phase flip}, \textit{bit flip}~\cite{nielsen2001quantum}. Furthermore, users can also customize a new noise themselves, or even choose a combination of all types of noise;
(ii) the type of quantum state, which can be either mixed or pure in the local component and is set as mixed by default in the global component. The choice for the global component is predetermined as global robustness verification for mixed states can be reduced to that for pure states; and
(iii) perturbation parameters, specifically two thresholds for robustness ($\epsilon,\delta$ in Problem~\ref{Prob:Global}) for the verification of \textit{global robustness} and a threshold for robustness ($\epsilon$ in Problem~\ref{Prob:local}) for the verification of \textit{local robustness}.

\subsection{Verifying Robustness}
\begin{figure}[h]
    \vspace{-0.4cm}
    \centering
    \includegraphics[width=\linewidth]{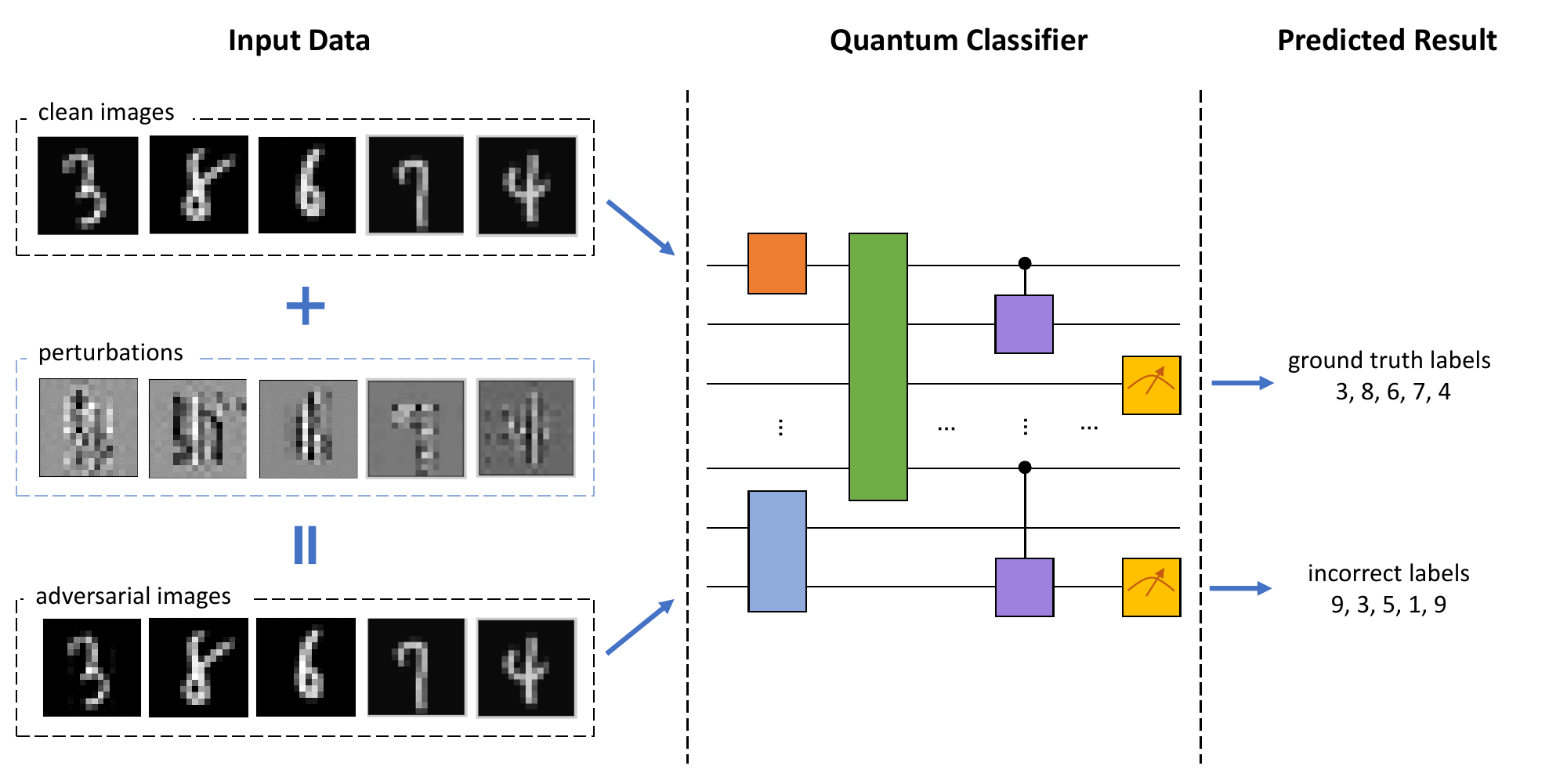}
    \caption{The adversarial examples and the corresponding adversarial perturbations found by \textit{VeriQR} in MINIST handwritten digit classification. }
    \label{fig:adv_example}
    \vspace{-0.8cm}
\end{figure}

\paragraph{{\textbf{Verification of local robustness.}}}
This part is comprised of five modules:
\begin{itemize}
    \item [1)]
    \textbf{Parser}: This module handles a quantum classifiers file to obtain the corresponding quantum circuit object. 
    \item [2)]
    \textbf{Noise generator}: The input for this module is a quantum circuit object. 
    \textit{First}, it generates a noisy quantum circuit by adding a random noise to each qubit at random points in the circuit with a randomly determined noise probability. The purpose of this is to simulate the effect of noise to verify the robustness of the QML model on real-world quantum hardware. 
    \textit{In addition}, users can also use \textit{VeriQR} to actively add noises of specific types, including commonly used standard quantum noise models and user-defined quantum noise models (using Kraus representation) with specific noise probabilities, to the noisy model. Here, user-specified noises are added at the end of the circuit, which is a common assumption. This functionality enables robustness improvements, illustrated by our experimental results in Section~\ref{sec:evaluation}. 
    \item [3)]
    \textbf{Constraint generator}: This module generates constraints based on the (noisy) quantum model and the input dataset, which are then submitted to the core verifier. 
    \item [4)]
    \textbf{Core verifier}: This module receives the constraints, a perturbation parameter $\varepsilon$, and the quantum state type as inputs. Based on the state type, it chooses the appropriate constraint solver: a Semidefinite Programming solver for mixed states, and a Quadratically Constrained Quadratic Program solver for pure states~\cite{guan2021robustness}. It then utilizes both the under-approximation and exact algorithms to initiate the verification analysis procedure for $\epsilon$-robustness. \emph{These algorithms and solvers are specifically designed to tackle the challenge of "Continuous State Space" discussed in Section~\ref{Sec:challenges} when verifying the robustness of QML models. In particular, the under-approximation algorithm is implemented to address the "State Explosion" issue.}
    \item [5)]
    \textbf{Statistics and visualization}: This module is responsible for visualizing and displaying the results in the GUI of \textit{VeriQR}. The computed robust accuracy of the quantum classifier, which indicates the validity of the robustness property, is outputted by \textit{VeriQR}. Additionally, the detected adversarial examples (quantum states) are stored in a NumPy data file for further analysis and to improve robustness through adversarial training. The GUI also presents the original and noisy quantum circuit diagrams for the classifier (see Fig.~\ref{fig:GUI}), providing users with an intuitive way to analyze the model construction. Moreover, for the MNIST handwritten digit classification task, \textit{VeriQR} displays pictures of the detected adversarial examples based on the digits specified by the user in the GUI. These adversarial examples are obtained by adding noisy perturbations to a set of legitimate input examples, as illustrated in Fig.~\ref{fig:adv_example}. 
\end{itemize}

{\textbf{Verification of global robustness.}} 
This part also includes five modules similar to validating \emph{local robustness}. However, instead of being verified directly by the verifier, \emph{the noisy model generated by the noise generator is first passed to the data structure converter and is transformed into the corresponding \emph{tensor networks} model to improve efficiency and overcome the challenge of "State Explosion" discussed in Section~\ref{Sec:challenges}. }
The core verifier then takes the tensor network model as input and calculates the required constant $K^*$ for model validation, following the procedure outlined in Algorithm 1 of \cite{guan2022verifying}. In addition to this, the core verifier receives perturbation parameters $\varepsilon$ and $\delta$ in Problem~\ref{Prob:Global} for validation and finally determines whether the global robustness property holds by checking if $\delta\geq K^*\epsilon$. If not, it will provide an adversarial kernel $(\psi, \phi)$, which is capable of generating infinitely many pairs of quantum states that violate the global robustness of the QML model. 

\subsection{Improving Robustness}\label{sec:improving_robustness}
\textit{VeriQR} offers adversarial training and adding specific noise to enhance the local and global robustness of QML models, respectively. 
The effectiveness of these strategies is validated through experiments on various QML models in Section~\ref{sec:evaluation}.

\vspace{-0.2cm}
\paragraph{{\textbf{Adversarial training.}}}
\textit{VeriQR} empowers users with adversarial training capabilities, an extension of traditional machine learning. When the $\epsilon$-local robustness of $\rho$ with label $l$ is compromised, our robustness verification algorithms embedded in \textit{VeriQR} automatically generate an adversarial example $\sigma$. By incorporating $(\sigma, l)$ into the training dataset, users can then retrain the QML model to enhance its local robustness against the adversarial examples.

\vspace{-0.2cm}
\paragraph{{\textbf{Specific noise.}}}
Previous research~\cite{du2021quantum,guan2022verifying,huang2023certified} suggests that introducing specific quantum noise at strategic points in the circuits of QML models can improve global robustness. The \textit{VeriQR} tool empowers users to apply standard or personalized noise at various locations in quantum circuits for robustness enhancement.


\section{Evaluation}\label{sec:evaluation}
In this section, we evaluate the effectiveness of \textit{VeriQR} in verifying and enhancing both the local and global robustness of various QML models. These models utilize popular parameterized quantum circuits like Quantum Neural Networks (QNN), Quantum Convolutional Neural Networks (QCNN), Quantum Approximate Optimization Algorithms (QAOA), Variational Quantum Eigensolver Algorithms (VQE) and Quantum Supremacy Algorithms. All these networks have previously demonstrated successful implementation on practical quantum hardware~\cite{zeguendry2023quantum}. All the experiments are performed on a workstation with a Intel(R) Xeon(R) Gold 6254 CPU @ 3.10GHz $\times$ 72 Cores Processor and 314 GB RAM.

\begin{table}[]
\vspace{-0.5cm} 
\centering
\small
\renewcommand\arraystretch{1.0}
\scalebox{.9}{
\begin{tabular}{p{1.0cm}<{\centering} p{1.5cm}<{\centering} p{1.0cm}<{\centering} p{1.2cm}<{\centering} p{3.0cm}<{\centering} cccc}
 \toprule
  \multirow{2}{*}{\textbf{Model}} & \multirow{2}{*}{\textbf{\#Qubits}} & \multirow{2}{*}{$\bm{\epsilon}$} & \multirow{2}{*}{\textbf{Circuit}} & \multirow{2}{*}{\makecell[c]{\textbf{Noise Setting} \\ ($noise\_p$)}} & \multicolumn{2}{c}{\textbf{Rough Verif}} & \multicolumn{2}{c}{\textbf{Accurate Verif}} \\ 
 \cmidrule(lr){6-7} \cmidrule(lr){8-9}
 & & & & & \textbf{RA (\%)} & \textbf{VT (s)} & \textbf{RA (\%)} & \textbf{VT (s)} \\
 \midrule
 
\multirow{4}{*}{\textit{qubit}} & \multirow{4}{*}{1} & \multirow{4}{*}{0.001} & $c_0$ & noiseless& 88.12 & 0.0038 & 90 & 2.4226 \\
 &  &  & $c_1$ & random & 88.12 & 0.0039 & 90 & 2.4623 \\
 &  &  & $c_2$ & depolarizing\_0.001 & 88.00 & 0.0038 & 90 & 2.4873  \\
 &  &  & $c_2$ & depolarizing\_0.005 & 87.62 & 0.0053 & 90 & 2.7140 \\
 \hline
 
\multirow{4}{*}{\textit{iris}} & \multirow{4}{*}{4} & \multirow{4}{*}{0.005} & $c_0$ & noiseless & 98.75 & 0.0013 & 100 & 0.4924 \\
 &  &  & $c_1$ & random & 97.50 & 0.0009 & 100 & 0.8876 \\
 &  &  & $c_2$ & mixed\_0.01 & 97.50 & 0.0019 & 100 & 0.8808 \\
 &  &  & $c_2$ & mixed\_0.05 & 96.25 & 0.0021 & 100 & 3.1675 \\
 \hline

\multirow{4}{*}{\textit{tfi}} & \multirow{4}{*}{4} & \multirow{4}{*}{0.005} & $c_0$ & noiseless & 86.41 & 0.0039 & 100 & 6.5220 \\
 &  &  & $c_1$ & random & 85.94 & 0.0038 & 100 & 6.6438\\
 &  &  & $c_2$ & mixed\_0.01 & 85.78 & 0.0061 & 100 & 6.7117 \\
 &  &  & $c_2$ & mixed\_0.05 & 85.16 & 0.0063 & 100 & 7.0374 \\
 \hline
 
\multirow{4}{*}{\textit{tfi}} & \multirow{4}{*}{8} & \multirow{4}{*}{0.005} & $c_0$ & noiseless & 98.44 & 0.0372 & 100 & 2.3004 \\
 &  &  & $c_1$ & random & 96.56 & 0.1061 & 100 & 3.9492 \\
 &  &  & $c_2$ & bit-flip\_0.01 & 96.56 & 37.0965 & 100 & 42.1246 \\
 &  &  & $c_2$ & bit-flip\_0.05 & 95.94 & 32.7195 & 100 & 38.8139 \\
 \hline
 
\multirow{4}{*}{\textit{fashion}} & \multirow{4}{*}{8} & \multirow{4}{*}{0.001} & $c_0$ & noiseless & 90.60 & 0.0420 & 97.40 & 25.3777 \\
 &  &  & $c_1$ & random & 90.30 & 0.0934 & 97.30 & 27.4964 \\
 &  &  & $c_2$ & bit-flip\_0.01 & 89.90 & 15.6579 & 97.20 & 42.1063 \\
 &  &  & $c_2$ & bit-flip\_0.05 & 87.60 & 14.0342 & 96.70 & 48.5805 \\
 \hline
 
\multirow{4}{*}{\makecell[c]{\textit{mnist} \\ $(1\&3)$}} & \multirow{4}{*}{8} & \multirow{4}{*}{0.003} & $c_0$ & noiseless & 93.80 & 0.0543 & 96.00 & 18.5063 \\
 &  &  & $c_1$ & random & 92.60 & 0.0785 & 95.70 & 23.2905 \\
 &  &  & $c_2$ & phase-flip\_0.001 & 92.60 & 12.9728 & 95.70 & 36.2348 \\
 &  &  & $c_2$ & phase-flip\_0.01 & 92.60 & 11.6704 & 95.70 & 33.7894 \\
 
 \bottomrule
\end{tabular}
}
\caption{Experimental results of the \textit{local robustness} verification of different QML models. }
\label{tab:local}
\vspace{-0.7cm} 
\end{table}

\subsection{Local Robustness} \label{sec:experiments_local}
We conducted several experiments to test the \textit{local robustness} of various quantum classifiers with different numbers of qubits. These classifiers were trained on labeled datasets that were encoded using different quantum encoders in platforms such as Mindspore Quantum~\cite{xu2024mindspore} and Tensorflow Quantum~\cite{broughton2020tensorflow}. The classifiers examined in our study include the \textit{qubit} classifier, which determines the qubit's position in the X-Z plane of a Bloch sphere~\cite{broughton2020tensorflow}; the \textit{iris} classifier, which categorizes irises from various subgenera~\cite{misc_iris_53}; the \textit{mnist} classifier, which identifies handwritten digits, specifically 1 \& 3~\cite{deng2012mnist}; the \textit{fashion} classifier, which classifies images of T-shirts and ankle boots~\cite{xiao2017fashion}; and the \textit{tfi} classifier, which recognizes wavefunctions at different phases in a quantum many-body system~\cite{broughton2020tensorflow}. 

\vspace{-0.2cm}
\paragraph{\textbf{Experiment setting for verification:}}
To investigate the impact of random and specific noise on local robustness verification, we conducted experiments on four different circuits for each model as outlined in Table~\ref{tab:local}: the noiseless ideal QML model with quantum circuit $c_0$; circuit $c_1$ created by introducing random noise at various random points in circuit $c_0$ to simulate noise effects on NISQ devices; and circuit $c_2$ modified by adding specific noise with a noise level $p$ (referred to as "noisename\_p" below $c_2$) of four types: \textit{depolarizing}, \textit{phase flip}, \textit{bit flip}, and \textit{mixed} (a combination of the three) noise, introduced in Section~\ref{subsection:QML}, applied to each qubit after the random noise manipulation on circuit $c_1$. 


\paragraph{\textbf{Experiment setting for approximate versus exact verification:}}
In each robustness verification scenario, we employed two verification techniques: a coarse method labeled "Rough Verif" and a precise method labeled "Accurate Verif". We must emphasize here the difference between accurate and rough verification methods for local robustness verification. The \textit{rough verification} method detects non-robust states only by applying the robust bound condition from the work in \cite{guan2021robustness}. However, quantum states that do not satisfy this condition may also be robust, leading to an underestimation of the robust accuracy. Therefore, the \textit{accurate verification} method first filters out possible non-robust states using the condition, and then uses a Semidefinite Programming solver to obtain the optimal robust bound for these states, thus verifying the local robustness of each state precisely. 

Table~\ref{tab:local} presents a summary of the outcomes obtained from our experiments on \emph{local robustness} verification. In this table, the robust accuracy of the classifiers is represented as "RA", while the verification time (in seconds) is indicated as "VT". The experimental results reveal two key aspects:

1. By examining the RA values in rows $c_0, c_1$, and $c_2$ for each QML experiment in Table~\ref{tab:local}, it becomes evident that both random noise and specific noise cannot enhance robustness, particularly in the \textit{fashion} and \textit{mnist} experiments.


2. When comparing the RA (VT) values between the "Rough Verif" and "Accurate Verif" columns, it is observed that the under-approximation of robust accuracy scales well in almost all cases with faster verification time, supporting the conclusions drawn in \cite{guan2021robustness}.


\begin{remark}
Our tool, \textit{VeriQR}, serves as a formal instrument capable of identifying all non-robust quantum states (adversarial examples) during the verification process of all quantum classifiers. Analogous to classical methodologies, adversarial training can be utilized to fortify non-robust states within the retrained models. In our study, we have incorporated the adversarial training technique as outlined in Section~\ref{Subsection:Robustness} for all QML models listed in Table~\ref{tab:local}. Despite being a conventional classical practice, we have documented the outcomes on our code repository.
\end{remark}

\begin{figure}[h]
    \vspace{-0.8cm} 
    \centering
    \subfloat{\includegraphics[width = 0.33\linewidth]{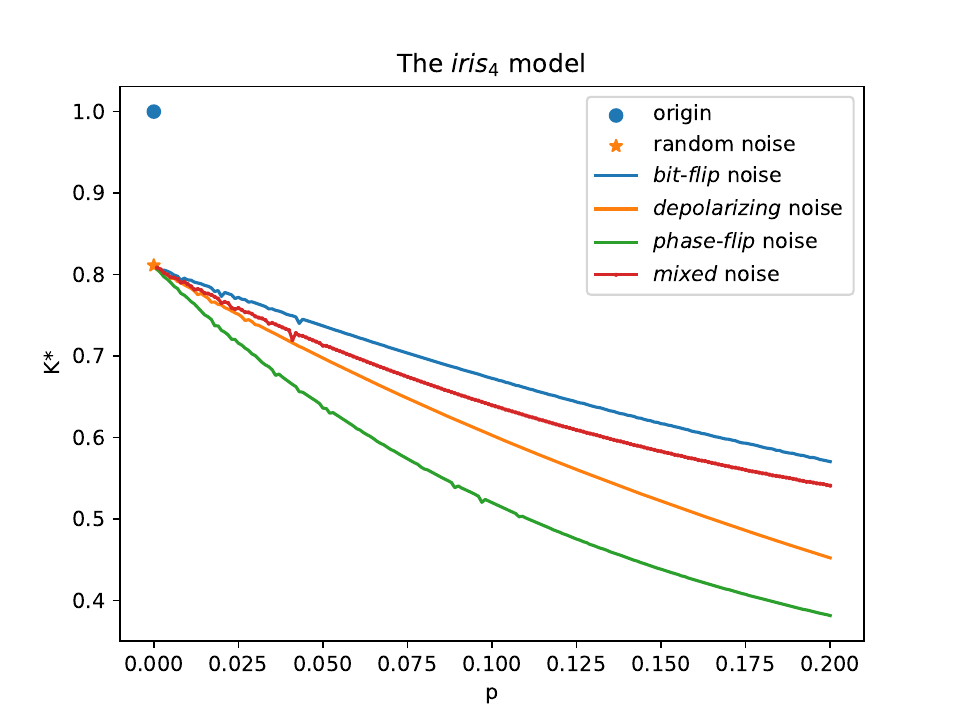}}
    \subfloat{\includegraphics[width = 0.33\linewidth]{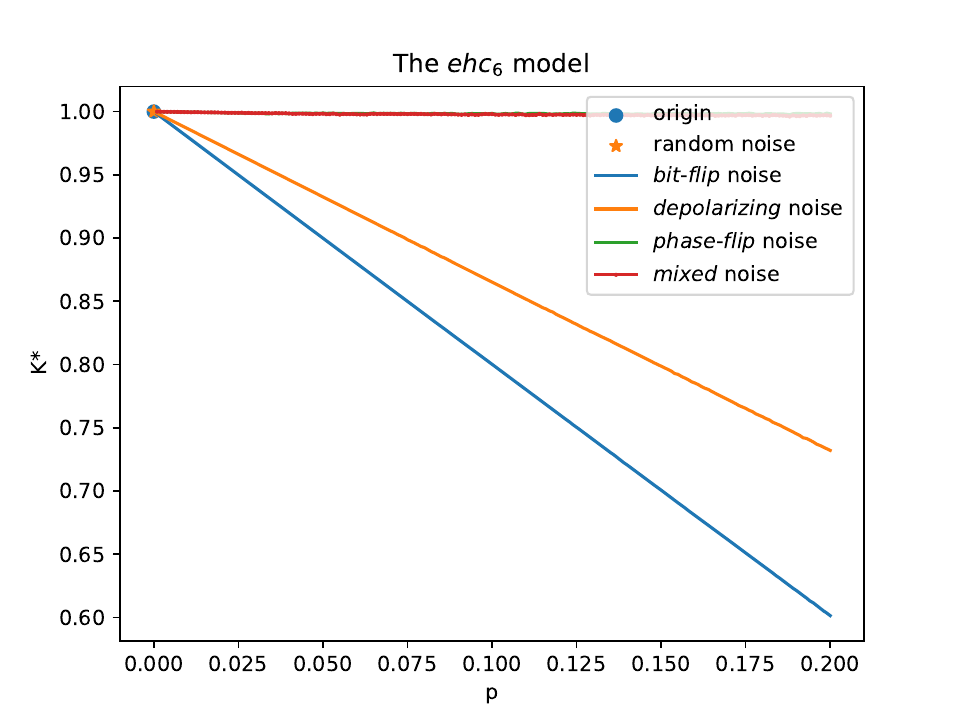}}
    \subfloat{\includegraphics[width = 0.33\linewidth]{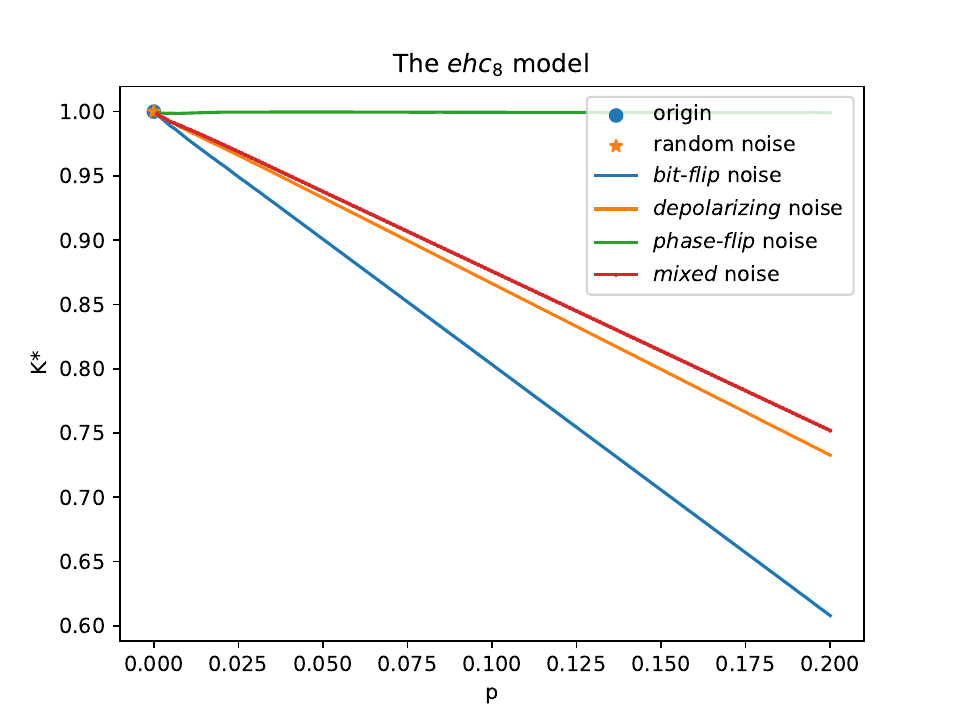}}
    \hfill
    \subfloat{\includegraphics[width = 0.33\linewidth]{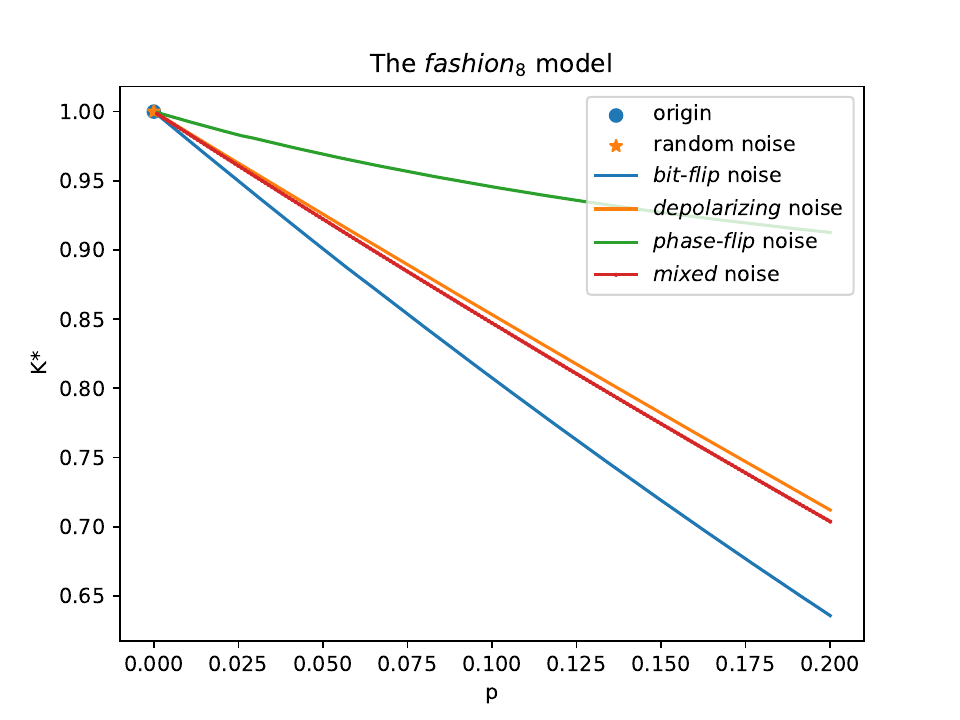}}
    \subfloat{\includegraphics[width = 0.33\linewidth]{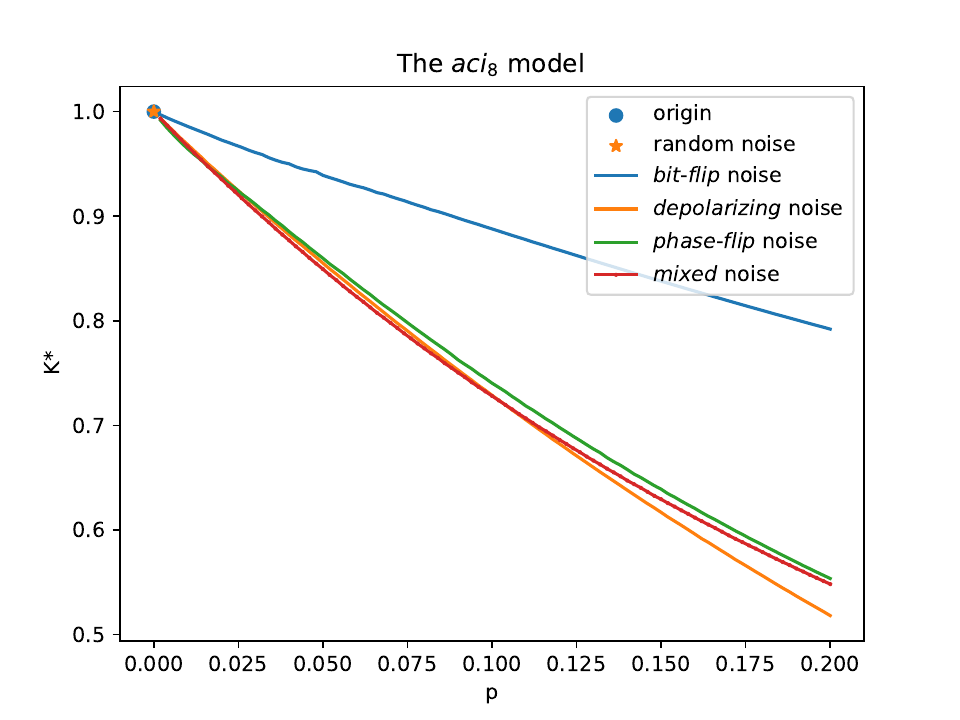}}
    \subfloat{\includegraphics[width = 0.33\linewidth]{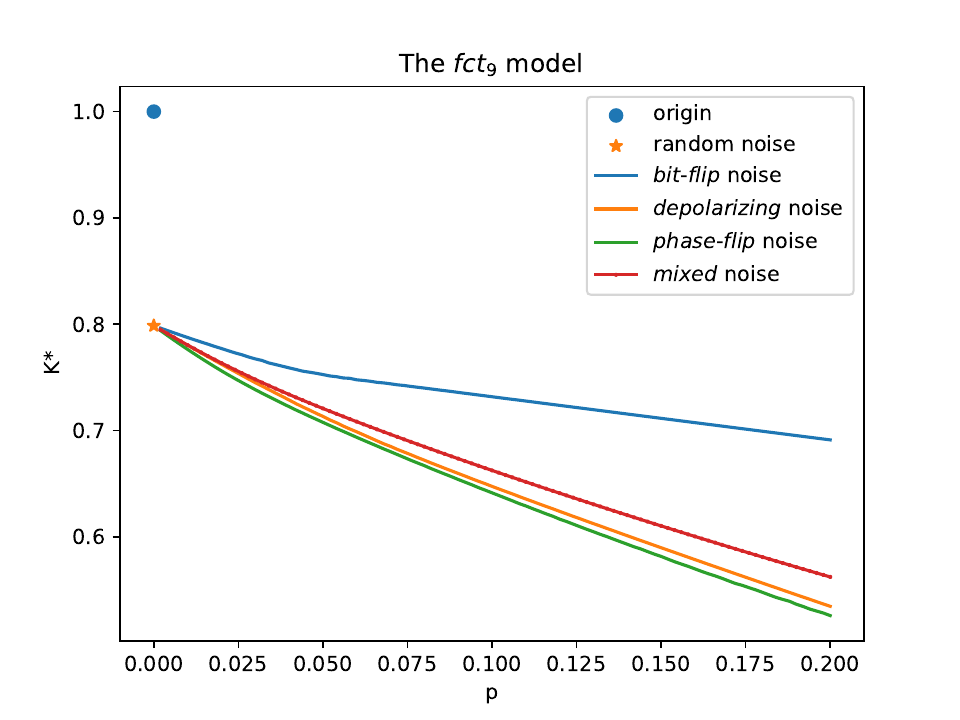}}
    \hfill
    \subfloat{\includegraphics[width = 0.33\linewidth]{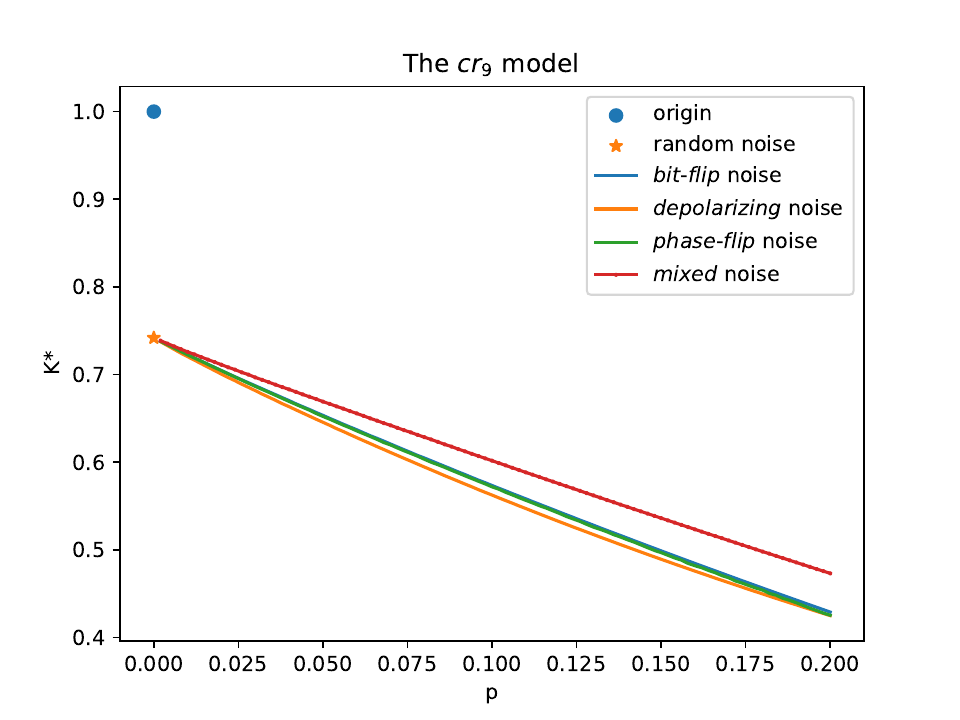}}
    \subfloat{\includegraphics[width = 0.33\linewidth]{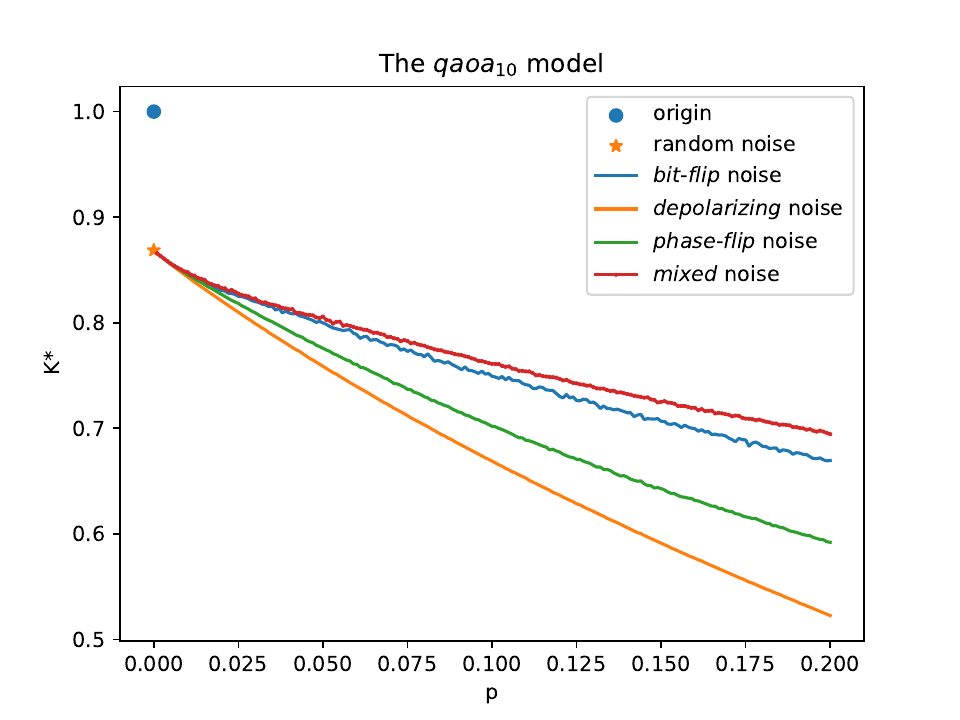}}
    \subfloat{\includegraphics[width = 0.33\linewidth]{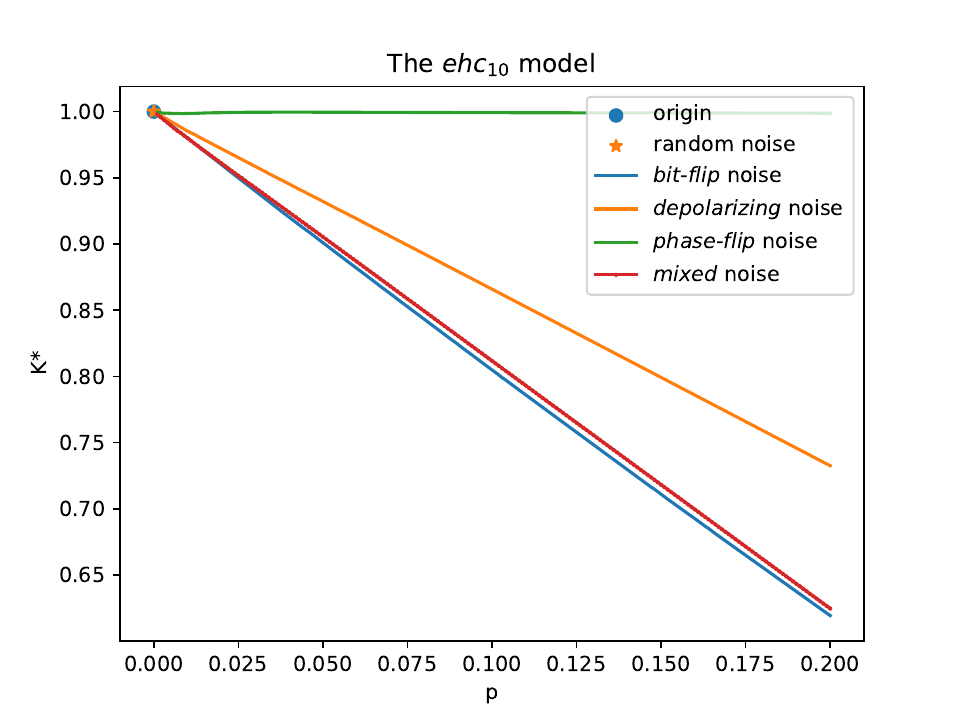}}
    \hfill
    \caption{Experimental results of the trade-off between the Lipschitz constant $K^*$ (measuring \textit{global robustness}) and noise level $p$ in different QML models. }
    \label{fig:global_result}
    \vspace{-0.7cm} 
\end{figure}

\subsection{Global Robustness}
For \textit{global robustness}, we also incorporate various types of noise and their corresponding noise levels into the quantum models to be tested. We conducted multiple experiments on different QML models using the \textit{VeriQR} tool. These experiments encompass a wide range of applications, including the \textit{aci} model for adult census income prediction \cite{ai_8}, the \textit{fct} model for detecting fraudulent credit card transactions \cite{ec_9}, the \textit{cr} model for classifying individuals as good or bad credit risks based on a set of attributes \cite{german_credit_data}, the \textit{ehc} model for calculating the binding energy of hydrogen chains \cite{google2020hartree}, the $qaoa$ model for solving hardware grid problems \cite{harrigan2021quantum}. 


\begin{table}[]
\centering
\small
\renewcommand\arraystretch{1.0}
\scalebox{.78}{
\begin{tabular}{p{1.3cm}<{\centering} p{1.3cm}<{\centering} p{2.3cm}<{\centering} c p{2.9cm}<{\centering} ccccc}
 \toprule
 \multirow{2}{*}{\textbf{Model}} & \multirow{2}{*}{\textbf{\#Qubits}} & \multirow{2}{*}{\textbf{Noise}} & \multirow{2}{*}{\bm{$p$}} & \multirow{2}{*}{\bm{$(\varepsilon, \delta)$}} & \multicolumn{2}{c}{\textbf{Baseline}} & \multicolumn{2}{c}{\textbf{TN}} & \multirow{2}{*}{\textbf{Robust}} \\ 
 \cmidrule(lr){6-7} \cmidrule(lr){8-9}
 & & & & & \bm{$K^*$} & \textbf{time (s)} & \bm{$K^*$} & \textbf{time (s)} & \\
 \midrule

 

\multirow{4}{*}{\textit{ehc}} & \multirow{4}{*}{8} & bit flip & 0.0001 & (0.0003, 0.0075) & 0.99980 & 0.26 & 0.99976 & 26.17 & YES \\
 &  & depolarizing & 0.05 & (0.001, 0.0075) & 0.93333 & 0.26 & 0.93304 & 27.87 & YES \\
 &  & phase flip & 0.025 & (0.075, 0.0003) & 1 & 0.26 & 0.99968 & 28.46 & NO \\
 &  & mixed & 0.0005 & (0.005, 0.005) & 0.99938 & 0.24 & 0.99905 & 25.75 & YES \\
\hline


\multirow{4}{*}{\textit{aci}} & \multirow{4}{*}{8} & bit flip & 0.0001 & (0.003, 0.0001) & 0.99985 & 0.18 & 0.99985 & 6.44 & NO \\
 &  & depolarizing & 0.025 & (0.03, 0.0005) & 0.92640 & 0.25 & 0.92440 & 7.70 & NO \\
 &  & phase flip & 0.05 & (0.05, 0.001) & 0.88450 & 0.19 & 0.85990 & 8.58 & NO \\
 &  & mixed & 0.005 & (0.005, 0.005) & 0.98384 & 0.22 & 0.98326 & 6.06 & YES \\
\hline

\multirow{4}{*}{\textit{fct}} & \multirow{4}{*}{9} & bit flip & 0.05 & (0.075, 0.003) & 0.99024 & 0.98 & 0.97683 & 13.89 & NO \\
 &  & depolarizing & 0.05 & (0.0003, 0.0001) & 0.92638 & 0.76 & 0.92486 & 40.73 & NO \\
 &  & phase flip & 0.01 & (0.01, 0.0075) & 0.98730 & 0.87 & 0.98290 & 10.45 & NO \\
 &  & mixed & 0.05 & (0.075, 0.0075) & 0.94531 & 0.89 & 0.92949 & 9.06 & NO \\
\hline

\multirow{4}{*}{\textit{cr}} & \multirow{4}{*}{9} & bit flip & 0.025 & (0.01, 0.0005) & 0.93964 & 0.65 & 0.93819 & 14.44 & NO \\
 &  & depolarizing & 0.005 & (0.075, 0.005) & 0.98637 & 1.21 & 0.98515 & 6.49 & NO \\
 &  & phase flip & 0.025 & (0.0003, 0.0001) & 0.94753 & 0.97 & 0.93772 & 9.63 & NO \\
 &  & mixed & 0.025 & (0.0001, 0.0001) & 0.95579 & 0.93 & 0.94980 & 12.15 & YES \\
\hline

\multirow{4}{*}{\textit{qaoa}} & \multirow{4}{*}{10} & bit flip & 0.005 & (0.05, 0.0005) & 0.99843 & 5.23 & 0.98507 & 16.98 & NO \\
 &  & depolarizing & 0.0001 & (0.01, 0.003) & 0.99983 & 6.15 & 0.99965 & 16.10 & NO \\
 &  & phase flip & 0.005 & (0.075, 0.0075) & 0.99224 & 5.14 & 0.98516 & 17.95 & NO \\
 &  & mixed & 0.001 & (0.03, 0.0075) & 0.99923 & 4.98 & 0.99657 & 16.16 & NO \\
\hline

\multirow{4}{*}{\textit{ehc}} & \multirow{4}{*}{10} & bit flip & 0.075 & (0.05, 0.0003) & 0.85409 & 3.37 & 0.85262 & 82.25 & NO \\
 &  & depolarizing & 0.0005 & (0.03, 0.001) & 0.99933 & 5.69 & 0.99924 & 40.33 & NO \\
 &  & phase flip & 0.01 & (0.0003, 0.0075) & 1 & 4.36 & 0.99857 & 66.67 & YES \\
 &  & mixed & 0.0001 & (0.005, 0.001) & 0.99981 & 5.26 & 0.99977 & 38.13 & NO \\
\hline

\multirow{4}{*}{\textit{ehc}} & \multirow{4}{*}{12} & bit flip & 0.005 & (0.0005, 0.0003) & 0.99001 & 169.42 & 0.98965 & 76.77 & NO \\
 &  & depolarizing & 0.0005 & (0.0001, 0.005) & 0.99933 & 253.11 & 0.99926 & 189.35 & YES \\
 &  & phase flip & 0.075 & (0.001, 0.0075) & 1 & 163.61 & 0.99880 & 675.50 & YES \\
 &  & mixed & 0.001 & (0.01, 0.0001) & 0.99997 & 195.48 & 0.99984 & 64.50 & NO \\
 \hline

 \multirow{4}{*}{\textit{inst}} & \multirow{4}{*}{16} & bit flip & 0.005 & (0.0005, 0.0003) & - & TO & 0.98009 & 1052.73 & NO \\
 &  & depolarizing & 0.0005 & (0.0003, 0.005) & - & TO & 0.99833 & 33.99 & YES \\
 &  & phase flip & 0.05 & (0.001, 0.0075) & - & TO & 0.95131 & 381.15 & YES \\
 &  & mixed & 0.001 & (0.005, 0.0003) & - & TO & 0.99899 & 123.25 & NO \\
 \hline

 \multirow{4}{*}{\textit{qaoa}} & \multirow{4}{*}{20} & bit flip & 0.05 & (0.005, 0.001) & - & TO & 0.91194 & 2402.32 & NO \\
 &  & depolarizing & 0.075 & (0.005, 0.003) & - & TO & 0.83488 & 433.05 & NO \\
 &  & phase flip & 0.0005 & (0.0001, 0.0001) & - & TO & 0.99868 & 70.00 & YES \\
 &  & mixed & 0.05 & (0.075, 0.0003) & - & TO & 0.89682 & 4635.55 & NO \\
 \bottomrule
\end{tabular}
}
\caption{Experimental comparison of tensor network-based verification with a baseline implementation lacking tensors for assessing \textit{global robustness}.}
\label{tab:global}
\vspace{-0.6cm} 
\end{table}

\vspace{-0.2cm}
\paragraph{{\textbf{Noise improving global robustness.}}}
Fig.~\ref{fig:global_result} depicts the scaling of the Lipschitz constant $K^*$ (which quantifies global robustness as discussed in Section~\ref{Subsection:Robustness}) across various models at different noise levels $p$ for four distinct noise types. The figure also showcases the experimental outcomes of the original model alongside a model derived from the original version with random noise. These results indicate that \emph{the global robustness of all models improves due to quantum noise}, as evidenced by the reduced $K^*$ value in the models. This outcome validates earlier theoretical findings suggesting that specific quantum noise can boost global robustness~\cite{du2021quantum,guan2022verifying,huang2023certified}.  The presence of "$\_n$" in each model name in the figure signifies the model's utilization of $n$ qubits.

\vspace{-0.2cm}
\paragraph{{\textbf{High efficiency of tensor network.}}}
Importantly, \textit{VeriQR} transformed quantum models into tensor network models and applied a tensor network-driven algorithm (referred to as "TN" in Table~\ref{tab:global}) for global robustness assessment. Table~\ref{tab:global} provides an experimental comparison with a baseline implementation (labeled as "Baseline") that does not incorporate tensors in global robustness evaluation. In this evaluation, a timeout threshold ("TO" entries) of 7,200 seconds was imposed. The results demonstrate that \emph{the tensor network approach significantly enhances verification speed for a large number of qubits (more than 12)}, thereby improving the scalability compared to the precise \emph{local robustness} verification outlined in Table~\ref{tab:local}.

\begin{remark}
    To further verify the robustness of \textit{VeriQR} both locally and globally, we have conducted additional experiments. These experiments involved testing the QML models presented in Table~\ref{tab:local} and~\ref{tab:global} but with varying numbers of qubits and different types and levels of noise. Furthermore, the experimental QML models encompass 45 MNIST classifiers that have been designed to classify all possible combinations of handwritten digits $\{0,1,2,\ldots,9\}$. All of these experiment results, along with the corresponding artifact for this paper, can be accessed in our code repository. 
\end{remark}

\section{Conclusion}
This paper presented \textit{VeriQR}, a graphical user interface (GUI) tool developed to verify the robustness of QML models in the current NISQ era, where noise is unavoidable. \textit{VeriQR} offers exact, under-approximate, and tensor network-based algorithms for local and global robustness verification of real-world QML models in the presence of quantum noise. Throughout the verification process, \textit{VeriQR} can identify quantum adversarial examples (states) and utilize them for adversarial training to improve the local robustness as the same as classical machine learning. Additionally, \textit{VeriQR} applies specific quantum noise to enhance the global robustness. Furthermore, \textit{VeriQR} is capable of accommodating any quantum model in the OpenQASM 2.0 format and can convert QML models into this format to establish a unified benchmark framework for robustness verification.

\begin{credits}
\subsubsection{\ackname}
We would like to thank Runhong He for his valuable discussion. This work was partly supported by the Youth Innovation Promotion Association,
Chinese Academy of Sciences (Grant No. 2023116), the Australian Research Council (Grant No. DP220102059), National Natural Science Foundation of China (Grants No. 62002333) and Innovation Program for Quantum Science and Technology (Grants No. 2021ZD0302901). This work was done when Yanling Lin was a remote research intern supervised by A/Prof. Ji Guan at the Institute of Software, Chinese Academy of Sciences.
\end{credits}

\bibliographystyle{splncs04}
\bibliography{references}

\end{document}